\documentclass[11pt,a4paper]{article}
\usepackage{latexsym}
\usepackage{amsfonts}
\usepackage{calrsfs}
\usepackage{amsmath}

\author{Diego Meschini\footnote{Corresponding author. E-mail address: diego.meschini@phys.jyu.fi} \qquad Markku Lehto \smallskip \\
\emph{Department of Physics, University of Jyv\"{a}skyl\"{a}, } \smallskip \\ \emph{PL 35 (YFL), FI--40014 Jyv\"{a}skyl\"{a}, Finland.}} 
\title{Is empty spacetime a physical thing?}
\date{October 24, 2005}
\begin{document}
\maketitle

\begin{abstract}
This article deals with empty spacetime and the question of its
physical reality. By ``empty spacetime'' we mean a 
collection of bare spacetime points, the remains of ridding spacetime 
of all matter and fields. We ask whether these geometric 
objects---themselves intrinsic to the concept of field---might 
be observable through some physical test. By taking quantum-mechanical notions into
account, we challenge the negative conclusion drawn from the
diffeomorphism invariance postulate of general relativity, and we
propose new foundational ideas regarding the possible
observation---as well as conceptual overthrow---of this geometric
ether.
\end{abstract}

\smallskip


\section{Introduction} \label{Int}
Since its inception, the ether has proved a troubled notion which,
contrary to common belief, still haunts physics at the turn of a
new century. What can one learn from the history of this ether and
in what sense is it still of utmost concern to physics?

The need for an ether as a material medium with mechanical
properties first became apparent to Descartes in the first half of
the seventeenth century in an attempt to avoid any actions that
would propagate, through nothing, from a distance. During its
history of roughly three centuries in its original conception, the idea
of the ether managed to materialize in endless forms via the works
of countless investigators. Its main purpose of providing a medium
through which interactions could propagate remained untouched, but
the actual properties with which it was endowed in order to
account for and unify an ever-increasing range of phenomena were
mutually incongruent and dissimilar. Never yielding to observation
and constantly confronted by gruelling difficulties, the
mechanical ether had to reinvent itself continually, but its very
notion staggered not a bit.

After almost 300 years of bitter struggle, the mechanical
ether eventually gave in. The first step of this change took place in the hands of Lorentz, for whom the ether was a sort of
substantial medium that affected bodies moving through it not
mechanically but only dynamically, i.e.\ due to the fact that
bodies moved through it. Drude and Larmor further declared
that the ether need not actually be substantial at all but simply
space with physical properties. The second and decisive step in
this de-mechanization of the ether was brought about by
Poincar\'{e} and Einstein through ideas that eventually took the shape of the special relativity theory. In Einstein's (1983) own words, this change ``consisted in taking away from the ether its last mechanical quality, namely, its immobility'' (p.~11).

Far from being dead, however, the ether had only transmuted its character---from a mechanical substance to an absolute inertial spacetime. The need for regarding inertial spacetime as an ether came after noticing that ``empty'' spacetime, despite being unobservable and unalterable, displayed physical properties, such as providing a reference for acceleration via its geodesics. 

The nature of this new ether underwent yet another change with the theory of general relativity. According to Einstein (1961, p.~176), the ether was now spacetime's dynamic and intrinsic metric content. This was so significant a change that it modified the very ideas of ether and empty spacetime. By making metric spacetime alterable, it actually put an end to its status as a genuine ether. And by making the metric field a content of spacetime, it did away with empty spacetime, since now to vacate spacetime means to be left with nothing at all. In fact, one \emph{is} left with something: the spacetime points; however, these had been denied physical reality by Einstein's hole argument. From this standpoint, Einstein concluded that empty spacetime cannot possess any physical properties, i.e.\ that empty spacetime does not exist.

It is the purpose of this article to challenge the certainty of
this conclusion. The way to achieve this will be connected with
the main ideas presented in our previous article (Meschini, Lehto, \& Piilonen, 2005), where the need to study the problem of the nature of empty spacetime (i) equipped with quantum theory and some guiding physical principle directly relevant to its existence---here proposed to be that of \emph{diffeomorphism invariance}---and (ii) from a non-geometric point of view was put forward. The first requirement stemmed from the simple fact that, without an appropriate guiding light, the search for new ideas becomes pure guesswork; the second originated from the observation that any genuinely new understanding of empty spacetime may require going beyond its geometric characterization entirely (further reasons justifying this second requirement will be given in this article). 
In particular both these observations were turned into criticisms of what is currently understood as pregeometry---there deemed a considerable incongruity.

When further heeding the history of the ether, the question of the
physical existence of empty spacetime must also involve that
\emph{observables intrinsic to empty spacetime} be found so as to be able to support any sound claim as to its reality. The present article constitutes the beginning of this endeavour.

\section{The mechanical ether}  \label{ME}
It is appropriate to start this investigation by tracing the rich
history of the ether, starting here from its older conception as a
material substance, passing through its virtual disappearance
after the progressive removal of its mechanical attributes, and
ending with its new form of an immaterial, geometric substratum,
as analyzed in the next two sections. For the historical review of this section, the very comprehensive
work of Whittaker (1951) will be followed as a
guideline.\footnote{Page numbers in parentheses in this section
are to Vol.~1 of this reference unless otherwise stated.}

Ren\'{e} Descartes (1596--1650) was the first to introduce the
conception of an ether as a mechanical medium. Given his belief
that action could only be transmitted by means of pressure and
impact, he considered that the effects at a distance between
bodies could only be explained by assuming the existence of a
medium filling up space---an ether. He gave thus a new meaning to
this name, which in its original Greek
($\alpha\grave{\iota}\theta\acute{\eta}\rho$) had meant the blue
sky or the upper air. The ether was unobservable and yet it was
needed to account for Descartes' mechanistic view of the universe,
given his said assumptions.\footnote{With his invention of the
coordinate system, Descartes was, at the same time, the unwitting
precursor of the later conception of the ether as a form of
space.} At the same time, the notion of an ether was right from
its inception entwined with considerations about the theory of
light. Descartes himself explained the propagation of light as a
transmission of pressure from a first type of matter to be
found in vortices around stars to a second type of matter,
that of which he believed the ether to be constituted. (pp.\ 5--9)

The history of the ether continued tied to the theory of light
with Robert Hooke's (1635--1703) work. In an improvement with
respect to Descartes' view, he conceived of light as a wave
motion, an exceedingly small vibration of luminous bodies that
propagated through a transparent and homogeneous ether in a spherical manner. Hooke also introduced thus the fruitful idea of a wave-front. (pp.\ 14--15)

Isaac Newton (1642--1727) rejected Hooke's wave theory of light on the grounds that it could not explain the rectilinear propagation of light or its polarization (see below). In its place, Newton proposed that light consisted of rays that interacted with the ether
to produce the phenomena of reflection, refraction and
diffraction, but that did not depend on it for their propagation.
He gave several options as to what the true nature of light might
be, one of which was that it consisted of
particles---a view that later on would be associated with his
name; nevertheless, as to the nature of light, he ``let every man
here take his fancy.'' Newton also considered it possible for the
ether to consist of different ``ethereal spirits,'' each
separately suited for the propagation of a different interaction.
(pp.\ 18--20)

Regarding gravitation in the context of his universal law of
attraction, Newton did not want to pronounce himself as to its
nature. He nonetheless conjectured that it would be absurd to
suppose that gravitational effects could propagate without the
mediation of an ether. However, Newton's eighteenth century followers
gave a twist to his views; antagonizing with Cartesians
due to their rejection of Newton's gravitational law, they went as
far as denying the existence of the ether---originally Descartes'
concept---and attempted to account for all contact interactions as
actions at a distance between particles. (pp.\ 30--31)

Christiaan Huygens (1629--1695) was also a supporter of the wave
theory of light after observing that light rays that cross each
other do not interact with one another as would be expected of
them if they were particles. Like Hooke, he also believed that
light consisted of waves propagating in an ether that penetrated
vacuum and matter alike. He managed to explain reflection and
refraction by means of the principle that carries his name, which
introduced the concept of a wave front as the emitter of secondary
waves. As to gravitation, Huygens' idea of a Cartesian ether led
him to account for it as a vortex around the Earth. (pp.\ 23--28)

An actual observation that would later have a bearing on the
notions of the nature of light and of the ether was that of
Huygens' regarding the polarization of light. He observed that
light refracted once through a so-called Icelandic crystal, when
refracted through a second such crystal, could or could not be seen depending on the orientation of the latter. Newton correctly understood this result as the first light ray being polarized, i.e.\ having properties dependent on the directions perpendicular to its direction of propagation. He then concluded that this was incompatible with light being a (longitudinal) wave, which could not carry such properties. (pp.\ 27--28)

Another thoroughly Cartesian account of the ether was presented by John Bernoulli (1710--1790), Jr., in an attempt to provide a
mechanical basis for his father's ideas on the refraction of
light. Bernoulli's ether consisted of tiny whirlpools and was interspersed with solid corpuscules that could never
astray much from their average locations. A source of light would temporarily condense the whirlpools nearest to it,
diminishing thus their centrifugal effects and displacing the said corpuscules; in this manner, a longitudinal wave would be started. (pp.\ 95--96)

In the midst of a general acceptance of the corpuscular theory of
light in the eighteenth century, also Leonhard Euler (1707--1783)
supported the view of an ether in connection with a wave theory of light after noticing that light could not consist of the emission of particles from a source since no diminution of mass was observed. Most remarkably, Euler suggested that, in fact, the same ether served as a medium for both electricity and light, hinting for the first time at a unification of these two phenomena. Finally, he also attempted to explain gravitation in terms of the ether, which he assumed to have more pressure the farther from the Earth, so that the resulting net balance of ether pressure on a body would push it towards the centre of the Earth. (pp.\ 97--99)

At the turn of the century, the wave theory of light received new
support in the hands of Thomas Young (1773--1829). Within this
theory, Young explained reflection and refraction in a more
natural manner than the corpuscular theory and, more importantly,
he also accounted successfully for the phenomena of Newton's rings (and hinted at the cause of diffraction) by introducing an
interference principle for light waves. It was Augustin Fresnel
(1788--1827) who, in 1816 and amidst an atmosphere of hostility
towards the wave theory, managed to explain diffraction in terms
of Huygens' and Young's earlier findings. (pp.\ 100--108)

Young and Fresnel also provided an alternative explanation of
stellar aberration, which had been first observed by James Bradley (1692--1762) in 1728 while searching to measure stellar parallax, and which had so far been explained in terms of the corpuscular theory of light. Young first proposed that such effect could be explained assuming that the Earth did not drag the ether with it, so that the Earth's motion with respect to it was the cause of aberration. Subsequently, Fresnel provided a fuller explanation that could also account for aberration being the same when observed through refractive media. Following Young, Fresnel suggested that material media partially dragged along the ether in such a way that the latter would pick a fraction $1-1/n^2$ (where $n$ is the medium's refractive index) of the medium's velocity. So far the ether was viewed as a somewhat non-viscous fluid that could be dragged along in the inside of refractive media in proportion to their refractive index, and whose longitudinal excitations described light. (pp.\ 108--113)

Considerations about the polarization of light would bring along
fundamental changes to the conception of the ether. As Newton had
previously observed, the properties of polarized light did not
favour a longitudinal-wave theory of light. Inspired by the
results of an experiment performed by Fran\c{c}ois Arago (1786--1853) and
Fresnel, Young hit on the solution to the problem of polarization
by proposing that light was a \emph{transverse} wave propagating in
a medium. Fresnel further hypothesized that the ether must then be
akin to a solid and display rigidity so as to sustain such waves. (pp.\ 114--117)

The fact that only a rigid ether could support transverse waves
robbed the idea of an immobile, undragged ether of much of its
plausibility, since it was hard to imagine a solid medium of some
sort that would not be, at least, partially dragged by bodies
moving through it. George Stokes (1819--1903) rose up to this challenge
by providing a picture of the ether as a medium that behaved like
a solid for high-frequency waves and as a fluid for slow moving
bodies. As a fluid, Stokes' ether was dragged along by
material bodies such that, in particular, it was at rest
relative to the Earth's surface. (pp.\ 128, 386--387)

Michael Faraday (1791--1867) gave a new dimension to the ether
conception by introducing the notion of \emph{field}, which in
hindsight was the most important concept to be invented in this
connection.\footnote{And this not without a sense of irony. This
is so because, at first, its was on the field, a physical entity
existing on its own and needing no medium to propagate, that the
overthrow of the mechanical ether would rest; however, later on
Einstein would reinstate the ether as a (metric) field
itself.} In his studies of the induction of currents, of the
relation of electricity and chemistry, and of polarization in
insulators, he put forward the concepts of magnetic and electric
lines of force permeating space. He introduced thus the concept of
a field as a stress in the ether and present where its effects
took place. He went on to suggest that an ether may not be needed
if one were to think of these lines of force---themselves part of
material bodies---as the carriers of transverse vibrations,
including light and radiant heat as well. Or then that, if there
was an luminiferous ether, it might also carry magnetic force and
``should have other uses than simply the conveyance of
radiations.'' By including also the magnetic field as being
carried by the ether, Faraday added to Euler's earlier prophecy,
and he hinted for the first time at the conception of light as an
electromagnetic wave. (pp.\ 170--197)

Another unifying association of this type was made by William Thomson (1824--1907), who in 1846 compared heat and electricity in that the isotherms of the former corresponded to the equipotentials of the latter. He suggested furthermore that electric and magnetic forces might propagate as elastic displacements in a solid. James Clerk Maxwell (1831--1879), inspired by Faraday's and W.\ Thomson's ideas, strove to make a mechanical picture of the electromagnetic field
by identifying static fields with displacements of the ether (for
him equivalent to displacements of material media) and currents
with their variations. At the same time, Maxwell, like Gustav
Kirchhoff (1824--1887) before him, was impressed by the equality of the measured velocity of light and that of
electromagnetic disturbances of his theory, and suggested that
light and electromagnetic waves must be waves of the same medium.
(pp.\ 242--254)

So far, the theories of Maxwell and Heinrich Hertz (1857--1894)
had not made any distinction between ether and matter, with the
former considered as totally carried along by the latter.
These theories were still in disagreement with Fresnel's
successful explanation of aberration in moving refractive media,
which postulated a partial ether drag by such bodies. However,
experiments to detect any motion of the Earth with respect to the
ether, such as those by Albert Michelson (1852--1931) and Edward
Morley (1838--1923), had been negative and lent support to Stokes' theory of an ether totally dragged at the surface of the Earth. (pp.\ 386--392)

Not content with Stokes' picture, in 1892 Hendrik Lorentz (1853--1928) proposed an alternative explanation with his theory of electrons, which reconciled electromagnetic theory with Fresnel's law. However, Lorentz's picture of the ether was that of an electron-populated medium whose parts were mutually at rest; Fresnel's partial drag was therefore not allowed by it. Lorentz's theory denied the ether mechanical properties and considered it only space with dynamic properties (i.e.\ affecting bodies because they moved through it), although still endowed it with a degree of
substantiality.\footnote{See (Kostro, 2000, p.~18) and (Kox, 1989, pp.~201, 207).} The negative results of the
Michelson-Morley experiment were then explained by Lorentz by
means of the existence of FitzGerald's contraction, which
consisted in a shortening of material bodies by a fraction
$v^2/2c^2$ of their lengths in the direction of motion
relative to the ether. Thus, the ether would cease being a
mechanical medium to become a sort of substantial, dynamic space.
(pp.\ 392--405)

Near the end of the nineteenth century, the conception of the
ether would complete the turn initiated by Lorentz with the views
of Paul Drude (1863--1906) and Joseph Larmor (1857--1942), which
entirely took away from the ether its substantiality. Drude (1894, p.~9) declared:
\begin{quote}
Just as one can attribute to a specific medium, which fills space
everywhere, the role of intermediary of the action of forces, one
could do without it and attribute to space itself those physical
characteristics which are now attributed to the ether.\footnote{Quoted from (Kostro, 2000, p.~20).}
\end{quote}
Also Larmor claimed that the ether should be conceived as an immaterial medium, not a mechanical one; that one should not attempt to explain the dynamic relations so far found in terms of
\begin{quote}
mechanical consequences of concealed structure in that medium;
we should rather rest satisfied with having attained to their
exact dynamical correlation, just as geometry explores or
correlates, without explaining, the descriptive and metric
properties of space.\footnote{Quoted from (Whittaker, 1951, Vol.~1, p.~303).}
\end{quote}
Larmor's statement is so remarkable that it will receive more
attention later on in Section \ref{BGE}.

Despite the seeming superfluousness of the ether even taken as a
fixed dynamic space, Lorentz held fast to the ether until his
death, hoping perhaps that motion relative to it could still
somehow be detected (Kox, 1989). Others, like Poincar\'{e} and Einstein, understood the repeated failed attempts to measure velocities with respect to the ether as a clear sign that the
ether, in fact, did not exist. Henri Poincar\'{e} (1854--1912) was the first to reach such a conclusion; in 1899 he asserted that absolute motion with respect to the ether was undetectable by any means, and that optical experiments depended only on the relative motions of bodies; in 1900 he openly distrusted the existence of the ether with the words ``Our ether, does it really exist?''; and in 1904 he proposed a principle of relativity. (Vol.~2, pp.~30--31) 

It was Albert Einstein (1879--1955) who in 1905 provided a theory where he reinstated these earlier claims but with a new, lucid
interpretational basis. In particular, Einstein considered
\begin{quote}
[T]he introduction of a ``luminiferous ether''\ldots to be
superfluous inasmuch as the view here to be developed will not
require an ``absolutely stationary space'' provided with special
properties\ldots (Einstein, 1952a, p.~38). 
\end{quote}
Thus, in the hands of Poincar\'{e} and Einstein, \emph{the ether had died}.

\section{Einstein's ethers} \label{EE}
Belief in the non-existence of the ether would, nevertheless, not
last very long, for it would soon rise from its
ashes---transmuted. A rebirth of the ether was now advocated by
Einstein on the grounds that, without it, ``empty'' space
could not have any physical properties; yet it displayed them
through the effects of absolute acceleration.

It is a well-known fact that all motion simply cannot be reduced to the symmetric relationship between any two reference
systems, as Newton's (1962, pp.~10--12) rotating-bucket and revolving-globes (thought) experiments, and Einstein's (1952b, pp.~112--113) rotating-spheres thought experiment aimed to show. Some reference frames clearly show their privileges. In Newtonian mechanics and in special relativity, these are the inertial frames; in general relativity, these are the freely-falling frames. 

Such directly unobservable frames of reference confer physical properties on ``empty'' spacetime, and were held by Einstein as a new embodiment of the ether. In order to understand Einstein's conceptions,\footnote{See (Kostro, 2000) for a useful source of material on Einstein and the ether. However, note that this book does not deal with the issue of the hole argument and the reality of spacetime points.} we now develop a further,
\emph{tentative} characterization of the ether by distinguishing
three different, possible realizations of it.

\begin{itemize}
\item[(i)] An ether is an entity that has sources and that cannot be observed directly, although it can be observed indirectly. This is the case of physical fields such as the metric, electric and magnetic
fields, $\mathbf{g}(x)$, $\vec E(x)$ and $\vec B(x)$, which can be
detected through the behaviour of test particles and influenced
through changes in their sources. We name these ethers
\emph{electric}, \emph{magnetic}, \emph{gravitational}, etc.
\emph{ether}, as the case might be.

\item[(ii)] An ether is an entity that does not have sources---therefore
cannot be influenced in any way---and cannot be observed directly,
although it can be observed indirectly via the behaviour of test
particles. Two constructs that realize this notion are Newtonian
space and special relativistic spacetime (characterized by the
constant metric field {\boldmath $\eta$}). We name these ethers \emph{inertial ethers}.

\item[(iii)] An ether is an entity that does not have sources---therefore cannot be acted upon in any way---and that ``acts'' but cannot be observed directly, nor indirectly through its effects on test particles. This kind of ether is realized by the \emph{spacetime points}, about which much will be said below and in Sections \ref{GE} and \ref{BGE}. For reasons that will become clear shortly, we name this ether a \emph{geometric ether}.
\end{itemize}

Einstein called Newtonian space and the metrics of special and
general relativity ethers. What reasons lie behind these identifications? 

The Newtonian and special relativistic inertial ethers above are ethers as genuine as the mechanical one had once been. By this we mean that, after allowing for a conceptual change from mechanical medium to geometric space, these inertial ethers are also directly unobservable substrata that nonetheless are thought to cause distinct effects in the observable world. While the mechanical ether---in its endless varieties---was regarded as the cause of effects ranging from gravitational to caloric, the inertial ethers afforded an explanation of motion.               

In effect, born out by his experiments according to which (the effect of) acceleration was not relative and rotation in empty space was meaningful, Newton held on to the need for a space absolutely at rest with respect to which this absolute acceleration could be properly defined.\footnote{In fact, only a family of inertial spaces linked by Galilean transformations would have sufficed judging from the properties of the theory itself.} Newton's absolute space is therefore an \mbox{``old new ether''} in sense (ii), since it has effects on test particles but has no sources and cannot be affected in any way.

From the point of view of the theory of special relativity, the
inertial ether is needed for the same reasons as above, but it
cannot be assigned any mechanical property whatsoever---not even
immobility or rest; any talk about its state of motion is
meaningless (Einstein, 1983, p.~13). This ether of special
relativity is nothing other than a background inertial spacetime, i.e.\ an infinite family of inertial frames linked by Lorentzian transformations. It is characterized by the everywhere-constant metric field {\boldmath $\eta$} and represents an indirectly observable ``empty'' spacetime endowed with physical properties: it defines the standards of space, time and motion for a test
particle in an otherwise empty world.

In spite of the historical success of the inertial ethers, we need not concern ourselves with them in this investigation. The reason is simply that, as independent concepts, they were overthrown by the theory of general relativity. Indeed, according to our best available understanding, the inertial ether is not an accurate description of Nature but only applies in certain limiting situations, and can be understood as a special case of the general relativistic metric $\mathbf{g}(x)$.     

General relativity conferred a totally new meaning on the concept of ether---so much so that the very notion of ether cannot be applied to metric spacetime any longer. Like before, the physical properties of spacetime are carried by its metric field---now $\mathbf{g}(x)$---which again defines the standards of space, time and motion. However, this metric field is a dynamic magnitude that is subject to change as dictated by the distribution of matter. In other words, the ether is no longer immutable but is revealed to have matter as its source, through which it can be acted upon.

Einstein had been explicitly concerned with this problem of finding an influenceable field to replace the prevailing immovable {\boldmath $\eta$}. When dealing with his rotating-spheres thought experiment, he said:
\begin{quote}
What is the reason for this difference [spherical and ellipsoidal] in the two bodies? No answer can be admitted as epistemologically satisfactory, unless the reason given is an \emph{observable fact of experience}\ldots [T]he privileged space $R_1$ [inertial] of Galileo\ldots is merely a \emph{factitious} cause, and not a thing that can be observed\ldots The cause must therefore lie \emph{outside} this system\ldots [T]he distant masses and their motions relative to $S_1$ and $S_2$ [the spheres] must then be regarded as the seat of the causes (which must be susceptible to observation) of the different behaviour of our two bodies $S_1$ and $S_2$. They take over the role of the factitious cause $R_1$. (Einstein, 1952b, pp.~112--113) 
\end{quote}
This shows that Einstein was looking for sources via which the
physical properties of spacetime could be influenced, and thus no longer fixed and beyond reach. In this sense, Einstein was the first physicist since the conception of the ether to bring it to full physical accountability (i.e.\ not only to passively observe its effects but also to control its structure) by finding, through the field equation, the ether's intimate linkage to matter as its source. He achieved this, however, not without altering the meaning of the original concept.

To be sure, this new property of $\mathbf{g}(x)$ greatly upsets its interpretation as an ether, since it puts it on virtually the same footing as other physical fields. Indeed, the metric field $\mathbf{g}(x)$ is akin to the electric field $\vec E(x)$ or magnetic field $\vec B(x)$ (or to the electromagnetic tensor field $\mathbf{F}(x)$). Whereas matter acts as the source of the former, charges act as sources of the latter; further, e.g.\ like seeing a compass move is evidence of the existence of $\vec B(x)$, so is seeing a stone fall evidence of $\mathbf{g}(x)$. The presence of all these fields can thus be granted beyond reasonable doubt.\footnote{And yet, this comparison is not as straightforward as one might wish. While electromagnetic fields carry physical units, the metric field does not, and neither does it result as the dimensionless quotient of other magnitudes. This makes the indirect observability of the metric field more intricate in comparison. It might be argued that one observes the effects of spacetime curvature and not really of the metric field.} 

Then, in order to understand Einstein's views regarding $\mathbf{g}(x)$ as an ether, we must recognize that the word can be meant not only in its later, negative sense of something physically unreal but also in its more primitive sense of an underlying, ubiquitous substratum with physical properties---even if \emph{now} it were on a par with other physical fields. While Einstein's usage highlights the all-pervading character of an ether, it blurs the more important issue of its physical reality or non-reality. As we saw above, as far as observability is concerned, the metric is a physical field analogous to the electromagnetic field; however, while the former necessarily permeates all of spacetime (a ubiquitous substratum), the latter does not. Like Einstein, also Weyl (1918, p.~182) recognized this point early on: 
\begin{quote}
The coefficients of the
fundamental metric form are therefore not simply the potentials of the gravitational and centrifugal forces, but \emph{determine in general which points of the universe are in reciprocal
interaction}. For this reason the name ``gravitational field'' is
perhaps too unilateral for the reality described by this
expression and should better be replaced by the word ``ether;''
while the electromagnetic field should simply be called field.\footnote{Quoted from (Kostro, 2000, p.~74).}
\end{quote} 
In order to also emphasize the aspect of the issue relating to physical existence, we reserve the use of the word ``ether'' in \emph{both} of the said senses, i.e.\ for denoting an underlying substratum that acts (has physical properties) but cannot be acted upon or observed through physical tests.

For one or more reasons given above, this description is not realized by the electromagnetic field, but neither by the general-relativistic metric. However, the description is realized by the spacetime points, since these geometric objects ``perform the localization of fields'' but they cannot be observed or influenced themselves as expressed by the diffeomorphism invariance postulate of general relativity (see Section \ref{GE}). Therefore, \emph{we call spacetime points the geometric ether}.\footnote{Notice that this denomination would be too inclusive if we had also accepted the metric field $\mathbf{g}(x)$ as an ether (or had concerned ourselves with {\boldmath $\eta$} as one), since its character is certainly also geometric. In any case, $\mathbf{g}(x)$ would have been a quantitatively geometric ether
(geometric magnitude), whereas spacetime points are only a qualitatively geometric ether (geometric object).}

Finally and in the same connection, not only did general relativity change the very notion of ether, but also that of empty spacetime. It is another feature of $\mathbf{g}(x)$ that it depends on the spacetime coordinates (geometrically speaking, on its points), so that it cannot consist of an absolute background associated with empty spacetime as {\boldmath $\eta$} in special relativity. On the contrary, the metric field $\mathbf{g}(x)$ constitutes an \emph{intrinsic content}\footnote{Einstein (1961) distinguished space from its contents with the words: ``[S]pace as opposed to `what fills space,' \emph{which is dependent on the coordinates}\ldots'' (p.~176) [Italics added]. This distinction makes sense since anything that depends on the spacetime coordinates (or points) must be \emph{in} spacetime.} of spacetime. Its removal means that the standards of space, time and motion (i.e.\ spacetime's geometric structure) are gone, so that not empty spacetime but, rather, nothing remains without it (Einstein, 1961, p.~176). Einstein's conclusion will be amplified in Section \ref{GE}, where the physical status of what \emph{does} remain---the spacetime points---will be investigated further.

\section{The geometric ether} \label{GE}
Einstein's conclusion that spacetime, empty of its metric-field
``ether,'' does not exist demands a qualification. This can be
seen as soon as one realizes that, after ridding spacetime of its
geometry, it is not ``nothing'' which remains but the spacetime points themselves. However, when Einstein (1961, 1983) seemingly jumped to conclusions in these (and other) expositions, he might have already been assuming as known the results of his so-called hole argument: spacetime points are not physical either, so that they could not constitute truly empty spacetime in any physical sense.

Given that Einstein's field equation is generally covariant, if
$\mathbf{g}(x)$ is a solution to it corresponding to the
matter-content source $\mathbf{T}(x)$, then so is
$\mathbf{g}'(x')$ with corresponding matter-content source
$\mathbf{T}'(x')$, for any continuous coordinate transformation $x
\mapsto x'$. This is simply so because the same matter and
metric fields are being viewed from two different frames of
reference, both being equivalent for the description of Nature.
Dropping the primed frame completely, how to express
$\mathbf{T}'(x')$ and $\mathbf{g}'(x')$ as viewed from the
unprimed frame? All that needs to be done is replace $x'$ by $x$
so that the matter and metric content are now seen from $S$ as
earlier seen from $S'$.

%\footnote{To see this with a simple example,
%imagine a function $x^2$ given with respect to coordinate system
%$S$ and a coordinate transformation $x'=x-a$ to a second
%coordinate system $S'$. The \emph{same} parabola from the point %of view of $S'$ is expressed as $(x'+a)^2$; this consists of a
%passive transformation. How to express the perspective of $S'$ %but only using the coordinate system $S$? Now an active %transformation must be performed on the original parabola $x^2$
%such that the space points are to be, so to speak, moved $a$ %units along the negative $x$ of $S$. The new function is %$(x+a)^2$, which represents the new view from $S$ of the %parabola as earlier seen from $S'$. Looking at the expression %$(x+a)^2$, it can be seen that it was obtained from the original %$(x'+a)^2$ by removing the primes. This is precisely the %procedure followed above, after Einstein's original, %coordinate-dependent way of thinking about these matters.}

This transformation consists of the active equivalent of a transformation of coordinates (from one reference frame to any another), and achieves its goal by ``moving'' the spacetime points and not the frame of reference. The result is that, since all frames of reference are physically equivalent, both $\mathbf{g}(x)$ with source $\mathbf{T}(x)$, and $\mathbf{g}'(x)$
with source $\mathbf{T}'(x)$ are solutions to the field equation
as seen from the same frame.

Now let $H$ be a so-called hole\footnote{The hole argument has
been reviewed in innumerable places. See e.g.\ (Stachel, 1989, pp.~71--81), where relevant quotations and a full list of early references can be found.} in spacetime in the sense that within this region no matter content is present, i.e.\ $\mathbf{T}(x)=0$; outside $H$, on the other hand, $\mathbf{T}(x)$ is non-null. Furthermore, let $\mathbf{g}(x)$ be the metric content of spacetime, necessarily non-null both inside and outside the hole, and $\phi$ an active
coordinate transformation with the property that it is equal to
the identity outside $H$ and different from the identity inside
$H$; demand also that the transformation be continuous at the
boundary of $H$. The property of general covariance of Einstein's
field equation now implies that the unchanged $\mathbf{g}(x)$,
with unchanged source $\mathbf{T}(x)$, is a solution outside $H$;
inside $H$, where no matter content is present, both
$\mathbf{g}(x)$ and $\mathbf{g}'(x)$ are (mathematically distinct)
solutions. Einstein's conclusion was that, their mathematical
differences notwithstanding, $\mathbf{g}(x)$ and $\mathbf{g}'(x)$
must be physically the same or else the field equation would not
be causal, since both metric fields are produced by the same
source outside $H$. The only way to avoid this dire consequence
was for Einstein to \emph{postulate the unreality of spacetime
points} since, in this manner, the above active transformation of
the metric field does not entail that there should be any
observational differences.\footnote{Notice that only the hole argument applied to \emph{general} relativity leads to the physical unreality of spacetime points. In special relativity, for example, the absolutely given metric field {\boldmath $\eta$} can be used to set up an inertial frame with respect to which spacetime points become physical events. In general relativity, on the other hand, there is no such possibility until the metric field $\mathbf{g}(x)$ has been obtained as a solution to Einstein's field equation; however, herein lies the problem: this field equation is not causal unless spacetime points are held to be physically unreal. See (Stachel, 1989, p.~78).}

There would be nothing remarkable about this conclusion if
spacetime points were otiose objects that contributed nothing to
our \emph{physical understanding}. Yet that is not the case. The
concept of field as a \emph{localized physical magnitude} (whether scalar, vectorial or tensorial), on which so much of our
scientific portrayal of the world (not just spacetime theories) is based, rests on the notion of point in order to have any meaning. In other words, points are intrinsic to the very concept of physical field.

In this respect, Auyang wrote:
\begin{quote}
The spatiotemporal structure is an integral aspect of the field\ldots We can theoretically abstract it and think about it while ignoring the dynamical aspect of the field, but our thinking does not create things of independent existence\ldots (Auyang, 2001, p.~214)
\end{quote}
He further criticized Earman's remarks that
\begin{quote}
When relativity theory banished the ether, the space-time manifold $M$ began to function as a kind of dematerialized ether needed to support fields. (Earman, 1989, p.~155) 
\end{quote}
We agree with Auyang's view that
``spacetime'' (spacetime points) is not merely a substratum on
which to \emph{mathematically} define fields. Spacetime points are inherent in fields inasmuch as they perform the \emph{physical}
task of localizing the latter. This is what, after Weyl (1949), Auyang (2001, p.~209) called the ``this'' or ``here-now'' aspect of a field, additional to its ``thus'' or qualitative aspect. In this view, spacetime points satisfy the traditional label of ``unobserved actors not acted upon,'' and hence our interpretation of points as an ether. However, although we also concur with Auyang in that points are the illusive creations of our geometric thinking, we do not offhand renounce the possibility that empty spacetime---\emph{beyond its geometric
description}---be real on its own, and not simply an illusion of our brains: might it not still be observable as a stand-alone entity?

A case in favour of the physical reality of spacetime points,
despite the odds against them, was put forward by Friedman (1983, pp.~216--263). According to him, the core of an explanation of natural phenomena is to be able to reduce a wide variety of them to a single framework (Friedman, 1974), so that what one is required to believe is not a vast range of isolated representational structures, but a single, all-encompassing construct. However, in order to provide scientific understanding, theoretical entities with \emph{sufficient unifying power} must be taken to be of a literal kind. Thus, Friedman argued that denying the existence of spacetime points---themselves essential for geodesics to exist---can only lead to a loss of unifying and explanatory power in spacetime theories.

As appealing as this argument may be, it is defeated by a look at history itself. If anything has been learnt from the narration of the history of the old ether, it should be this: no other conception gripped the minds of so many illustrious men of science for longer and more strongly than that of the ether. It was always held to be physically real and unified a wide range of phenomena (light, heat, gravitation, electricity and magnetism) despite its relentlessly unobservable existence. And yet, at the opening of the twentieth century, it became superfluous and useless, and was declared nonexistent.

Evidently, no spell of time during which a conception proves
to be extremely successful is long enough to declare it real because of
its utility or unifying power. What is to guarantee that today's
extremely successful spacetime points---our geometric ether---will
not run, in their own due time, the same fate as their mechanical
ancestor? If any concept that is of aid in physics is to be
held real, nothing more and nothing less should be demanded of it
that it be \emph{observable}.

Now, in studying current theoretical pictures comprised under the
generic label of quantum gravity (including pregeometry), one can
notice that their overall trend is not to try to overcome the
geometric ether as here explained by attempts to observe
\emph{empty} spacetime, but to create ever more involved metric
(i.e.\ non-empty) instantiations of it and then, possibly, to
observe those. In other words, quantum gravity has come to be an
attempt to replace general relativity's quantitatively geometric
picture of spacetime by other equally quantitatively geometric
pictures which, at the same time, include unobservable geometric
objects.

In this respect, Brans (1999, pp.~597--602) argued that much by way of directly unobservable structure is taken for granted in current spacetime theories, such as the existence of a point set, a topology, smoothness, a metric, etc. He compared the vortices
and atoms of the mechanical ether with these unobservable building blocks of spacetime and, moreover, with the yet considerably more complex spacetime structures and ``superstructures'' devised more recently. We interpret Brans to be asking: will strings and membranes, spin networks and foams, nodes and links---to name but a few---appear a hundred years from now like the mechanical ether does today? Are the above the new geometric counterparts of the old mechanical contraptions?

Be this as it may, even if these more recently
proposed structures were not to follow such a fate and could
eventually be observed, the observations relating to
them would give evidence of a new quantitatively geometric
constitution of spacetime---therefore, of a \emph{filled}
spacetime.\footnote{Hypothetical observation of the effects of a spacetime lattice (e.g.\ Smolin, 2004, p.~64) could perhaps be argued as indirect evidence of the vertices themselves, but certainly also as the effects of the lattice's full metric structure.} However, if one's endeavour is to advance the
understanding of \emph{empty} spacetime, then one must move away
from matter and geometry---forwards on a new path.

\section{Beyond the geometric ether} \label{BGE}
Is Einstein's conclusion (Section \ref{GE}) that there is nothing
beyond the gravitational ``ether'' $\mathbf{g}(x)$ final, then? It
is not as long as an issue remains. Might it not be possible to
observe this ``nothing,'' the geometric ether? Can one find
observables intrinsic to empty spacetime itself?

The story of the mechanical ether is repeating itself today in a
geometric, instead of mechanical, guise. The old ether was superseded by stripping it of all its intrinsic,
mechanical properties and rendering them superfluous; the old
ether did not exist. The new ether is also unobservable as the old one and as problematic, in the sense that it hints at the presence of an unresolved physical issue: from where observable does empty spacetime's physical capacity to ``localize fields'' come? To answer this question, the new ether must, like its ancestor, have
another layer of its nature revealed, again by stripping it of its
intrinsic properties: this time, geometric ones. Our belief is
that, having gone beyond the geometric ether, i.e.\ beyond geometry
entirely, observables intrinsic to empty spacetime might be
identified.

To put it differently, we are here proposing an \emph{updated}
version of Larmor's centenary words:
\begin{quote}
We should not be tempted towards explaining the simple group of
relations which have been found to define the activity of the
aether by treating them as mechanical consequences of concealed
structure in that medium; we should rather rest satisfied with
having attained to their exact dynamical correlation, just as
geometry explores or correlates, without explaining, the
descriptive and metric properties of space.\footnote{Quoted from (Whittaker, 1951, Vol.~1, p.~303).}
\end{quote}
Larmor's statement is remarkable for its correctness, and all the
more remarkable for its incorrectness. Its first half (up to the
semicolon) is a correct testimony of what soon would prove itself
the way out of the mechanical ether problem: its denial by special
relativity. Its second half \emph{was}---at the time of its
utterance---a correct comparison between the way Larmor thought
the mechanical ether should be considered and the way geometry was
then regarded, namely, not as background for the explanation of
phenomena. However, 16 years later in 1916 the ether would be
overtly\footnote{In fact, Newton's space had already assumed this
role over 200 hundred years before, and special relativity's
inertial spacetime 11 years before (see Section \ref{EE}), but
neither of them had been openly considered as spaces with physical geometric properties until after 1916.} reinstated as a dynamic field, and the physical properties of spacetime would be \emph{explained} by it in the sense of a geometric substratum. This trend of attempting to explain phenomena in terms of geometry did not stop with general relativity but continued with the efforts of quantum gravity and, in particular, of so-called pregeometry. Therefore, nowadays geometry does explain, as a substratum, the properties of spacetime, and the second part of Larmor's view is no longer correct.

The updated version of Larmor's words proposed here reads
thus---We should not be tempted towards explaining the simple
group of relations (\emph{fields' local aspect})
which have been found to define the activity of the
\emph{geometric ether} (\emph{spacetime points}) by treating them
as consequences of concealed structure in that \emph{geometric
medium}; \emph{we should rather seek to explain the activity of
the geometric ether beyond its geometric nature, searching
for observables intrinsic to empty spacetime via non-geometric
concepts}.

Unlike Larmor, we do not renounce an explanation of the geometric ether,
but we do not attempt to find it at the same conceptual level as
this ether finds itself.

Our search for spacetime observables requires taking a
\emph{conceptual} step beyond the state of affairs as left by
Einstein's hole argument. We believe that the current
philosophical literature on this problem (e.g.\ Butterfield, 1989; Earman \& Norton, 1987; Rynasiewicz, 1994, 1996)
has not been able to take this step by adding something
\emph{physically} new to the discussion. On the contrary, the
philosophical debate appears to function in the spirit of Earman's (1989) words, which measure the fruitfulness of a work by asking ``How many discussions does it engender?'' (p.~xi). What is needed is a new physical insight by means of which the present philosophical debate may be rendered inconsequential, much like the older disputes as to the shape and position of the Earth or the nature of the heavenly bodies were only settled by new physical investigations.

In order to understand how our step beyond the hole argument is
related to the hole argument itself, we will now re-rehearse it
but this time in geometric, rather than coordinate, language and
for a general field. Let there be two points, $P$ and
$Q$, inside $H$ linked by a diffeomorphism $\phi(P)=Q$, and let each of them
be the local aspect, or location, of fields $f(P)$ and $f'(Q)$,
respectively, solutions to the field equation with the same
source. The demand that
\begin{equation}
f(P)=f'(Q) \label{coordinate}
\end{equation}
(cf.\ $\mathbf{g}(x)=\mathbf{g}'(x')$) reproduces geometrically
the requirement that, after a coordinate transformation $x\mapsto
x'$, a physical situation remains unchanged. In terms of an active
transformation, point $P$ (cf.\ $x$) is dropped, and $f(P)$ is now
represented only in terms of $Q$ (cf.\ $x'$) by its push-forward
\begin{equation}
\phi^\ast [f(P)]=f[\phi(P)]=f(Q). \label{active}
\end{equation}
This is also a solution, but
\begin{equation}
f(Q)\neq f'(Q) \label{different}
\end{equation}
(cf.\ $\mathbf{g}(x')\neq \mathbf{g}'(x')$).\footnote{It would
have been more natural to denote $Q$ by $P'$ and later to perform
an active transformation dropping $P'$ (cf.\ $x'$) instead of $P$
(cf.\ $x$), in which case one would have obtained the expressions
$f(P)=f'(P')$ instead of (\ref{coordinate}) and $f(P)\neq f'(P)$
instead of (\ref{different}), more in tune with Einstein's
original coordinate notation. However, remaining faithful to this
starting point would have somewhat obscured the notation in the
investigation that follows at the end of this section.} Again, in
order to preserve the causality of the field equation, one
postulates diffeomorphism invariance, i.e.\ that ``displacements'' of points lead to no observable effects and that, therefore, points have no physical reality.

How to move forward, then? We note that the meaning of the
postulate stated above is only at face value so. On closer
inspection, diffeomorphism invariance involves only the weaker
requirement that points be all alike, i.e.\ have no physical
identity; thereby, the problem posed by the hole argument is
avoided, since having physically indistinguishable spacetime points
makes $\mathbf{g}(P)$ and $\phi^\ast[\mathbf{g}(P)]$ physically
equivalent.

In order to appreciate why having physically indistinguishable points solves the hole argument, one must clearly recognize the difference between mathematical and physical points. Mathematical points may very well be labelled points but, as Stachel remarked,
\begin{quote}
[N]o mathematical coordinate system is \emph{physically} distinguished per se; and without such a distinction there is no justification for physically identifying the points of a [mathematical] manifold\ldots as physical events in space-time. Thus, the mathematician will always correctly regard the original and the dragged-along fields as distinct from each other. But the physicist must examine this question in a different light\ldots (Stachel, 1989, p.~75) 
\end{quote}
The physicist must, in this case, rather ask whether there is anything in \emph{empty} spacetime by means of which its points can be physically told apart from one another---neither labels nor metrics can count to carry out the differentiation. Finding there are no such means, the physicist must hold spacetime points physically identical to one another. 

However, having a multitude of physically indistinguishable spacetime points does not necessarily mean that they must be physically meaningless in every other way (cf.\ hydrogen atoms; see below). 

An earlier attempt by one of us (Lehto, Nielsen, \& Ninomiya, 1986a,b) in which the principle of diffeomorphism invariance and quantum theory were both taken into account, revealed that fields on a pregeometric lattice displayed quantum-mechanical \emph{correlations}. In that work, the quantum-mechanical framework was realized via a path-integral formalism; within it, the requirement of diffeomorphism invariance demanded that one ought to sum over all the vertices $n$ in the partition function $Z$, thus leading to the appearance of correlations. One the other hand, the said requirement induced a free gas behaviour\footnote{Free gas behaviour in this case means that any pair of vertices can with high probability have any mutual distance.} of the vertices, which helped to avoid the rise of long-range correlations.\footnote{The correlation function 
$|\langle f(P)f(Q)\rangle - \langle f(P)\rangle \langle f(Q)\rangle |$
for fields on the lattice was found to be non-null, although correlations were nevertheless semi-local, i.e.\ actually tending to zero exponentially as the distance between two lattice points $P$ and $Q$ increased.}

Although we do not plan to follow this earlier approach, we rescue from it the insinuation that the said principle and quantum theory---the only branch of natural science so far forced to confront the problem of \emph{existence}\footnote{See (Isham, 1995, p.~65).}---can lead together to the result that physical fields can be mutually correlated in spacetime.
We hold such correlations to be the key to the possible identification of spacetime observables.

An analogy  with matter that behaves quantum-mechanically may
clarify in what sense correlations can fulfill this task. Just like spacetime points are physically identical, so are the hydrogen atoms conforming a gas of this element. However, denying the reality of hydrogen atoms on these grounds does not seem at all reasonable (Horwich, 1978, p.~409; Friedman, 1983, p.~241). Hydrogen atoms, despite being identical, possess a property that spacetime points do not seem to have: they interact with other matter and \emph{correlate} with one another creating bonds, hydrogen molecules. Holding the hydrogen molecules to be our sought-for observables for the sake of this analogy,\footnote{In a genuinely empty spacetime, points could only ``interact'' among themselves. Therefore, the interaction of the hydrogen atoms with e.g.\ their container's walls is an analogy we cannot pursue.} their measurable properties now give evidence of the existence of the physically identical atoms.

Forty years ago, Wheeler touched upon an idea somewhat similar to ours within the context of his so-called ``bucket of dust.'' He wrote:
\begin{quote}
Two points between which any journey was previously very long have
suddenly found themselves very close together. However sudden the
change is in classical theory, in quantum theory there is a
probability amplitude function which falls off in the classically
forbidden domain. In other words, there is some residual
connection between points which are ostensibly very far apart.
(Wheeler, 1964, p.~498) 
\end{quote}
However, Wheeler did not develop this idea, but only used it in order to reject his quantum-mechanical concept of nearest neighbour, according to the manner in which he had previously defined it.

In order to display the physically new idea of spacetime correlations within the existing context of the hole argument---but moving, at the same time, beyond it---we proceed as follows. In an otherwise empty spacetime, a field $f(P)$ at point $P$ dragged onto another point $Q$,
\begin{equation}
\phi^\ast [f(P)]=f[\phi(P)]=f(Q),
\end{equation}
and \emph{then pulled back again}, $\phi_\ast [f(Q)]$, could carry properties pertaining to $Q$ back with it, so that a comparison of $f(P)$ and $\phi_\ast [f(Q)]$ could yield that they are physically different. 

This is not to say that we expect to find new physics by means of a purely mathematical operation plus its inverse. It rather means that, given the insight that indistinguishable points may nonetheless display physical effects by correlating with each other quantum-mechanically, then the above analysis constitutes a means to represent the ensuing broken physical symmetry of diffeomorphism invariance. Moreover, when we speak of ``moving points,'' no physical system is actually being displaced in the physical world; the physical meaning of the expressions above (and below) ultimately falls back on spacetime correlations, the existence of which is here conjectured.

Given $f(P)$, the appearance of the pull-back $\phi_\ast [f(Q)]$ rests on the need to compare locally the original field and the original field restored. This need arises from the fact that physical experiments are not performed globally but locally. An analogy close at hand is that of parallel transport in general relativity. Given two vector fields $\vec v(P)$ and $\vec v(Q)$ in curved spacetime, they can only be compared by computing $\Delta \vec v$ at one point $P$ by parallel-transporting the latter field: 
\begin{equation}
\Delta \vec v(P)=\vec v(Q)_\parallel - \vec v(P).
\end{equation} 
The analogy works best for the case of a small, closed loop with sides $\Delta a \vec e_1$ and $\Delta b \vec e_2$. In this case, the field $\vec v(P)$ is compared to itself after having travelled the loop and returned to its original position; 
\begin{equation}
\Delta \vec v(P)=\Delta a  \Delta b \mathbf{R}(\ \ ,  \vec v(P), \vec e_1, \vec e_2), 
\end{equation}
where $\mathbf{R}$ is the Riemann tensor, quantifies how much the components---but not the size---of $\vec v(P)$ have been changed by the experience. Similarly, if instead of thinking of a field as a mere value, we visualize it as a \emph{vector} $f$, and $f(P)$ as a \emph{component} of $f$, then a new dimension to the hole argument is revealed: the local comparison of $\phi_\ast [f(Q)]$ and $f(P)$ quantifies how much the field component $f(P)$---but not the field's value---changes due to the possible correlation of $P$ with $Q$; in short, $f$ as a vector would behave like $\vec v$. This change would stem from any physical reality of quantum-mechanical origin that the points may have; pictorially speaking, $P$ would behave as if it ``remembered'' having ``interacted'' with $Q$, remaining ``entangled'' with it.

Now in the same manner that the parallel-transport procedure constitutes the mathematical representation of a physical property of spacetime---namely, its curva\-ture---so does our description intend to represent a \emph{new} kind of physical property---namely, spacetime correlations. Further, just as curvature reveals itself as the active effect of spacetime's geometric structure and may well be viewed as a sort of lingering connection between the points on the loop, the correlations we envision would likewise unveil themselves as the active effect of a persistent connection between spacetime points. However, in this case, we must search for the source of this connection somewhere else, within a deeper layer of the nature of things (see below).

The challenge now is to find \emph{observables} which reveal this possible behaviour of field $f(P)$, since fields alone are not observable as such. In this connection, we note with Einstein
(1952b, pp.~119, 121, 131; 1970, p.~71; 1982, p.~47) that the intervals $\mathrm{d}s^2$---and not the metric field---are the fundamental constituents of general relativity: the theory is
essentially about an \emph{observable} network of invariant
intervals between events. Through the intervals, a featureless
spacetime acquires geometric structure, which \emph{then} can be
characterized via the metric tensor (and the inner product), thus:
\begin{eqnarray}
\mathrm{d}s^2= \mathrm{d}\vec s \cdot \mathrm{d}\vec s & = &\left( \frac{\partial \vec s}{\partial x^\mu}\Big\vert_P\mathrm{d}x^\mu\right)\cdot \left( \frac{\partial \vec s}{\partial x^\nu}\Big\vert_P\mathrm{d}x^\nu\right) \nonumber \\
&=& \left( \frac{\partial \vec s}{\partial x^\mu}\Big\vert_P\cdot \frac{\partial \vec s}{\partial x^\nu}\Big\vert_P \right)\mathrm{d}x^\mu \mathrm{d}x^\nu \nonumber \\
&=&g_{\mu\nu}(P) \mathrm{d}x^\mu \mathrm{d}x^\nu.
\end{eqnarray}

In keeping with this crucial realization about $\mathrm{d}s^2$, 
the interval between points suggests itself as the main candidate for a spacetime observable. One could explore\footnote{Here $P'$ and $Q'$ are points in the neighbourhoods of $P$ and $Q$ respectively, and are
linked by a diffeomorphism $\phi(P')=Q'$, in the same way that $P$ and $Q$ are.} whether the interval
\begin{equation}
\mathrm{d}s^2_{PP'}=g_{\mu\nu}(P)\mathrm{d}x^\mu \mathrm{d}x^\nu \label{intial}
\end{equation}
between two points\footnote{Special care must be taken here not to confuse points with events.} remains unchanged after $g_{\mu\nu}(P)$ is pushed forward,
$\phi^\ast [g_{\mu\nu}(P)]$, to get
\begin{eqnarray}
\mathrm{d}s^2_{QQ'}&=&\phi^\ast [g_{\mu\nu}(P)]\mathrm{d}x^\mu
\mathrm{d}x^\nu \nonumber\\
&=&g_{\mu\nu}(Q)\mathrm{d}x^\mu \mathrm{d}x^\nu, \label{dragged}
\end{eqnarray}
and subsequently pulled back, $\phi_\ast [g_{\mu\nu}(Q)]$, to get
for the original interval
\begin{equation}
\phi_\ast [g_{\mu\nu}(Q)] \mathrm{d}x^\mu \mathrm{d}x^\nu. \label{restored}
\end{equation}
If spacetime points had any physical reality, the final expression (\ref{restored}) could not be equal to the
initial one (\ref{intial}), and there would be some long-range
correlations seen in the line element. 

This would not upset general relativity because, as said above, the envisioned correlations represent a new kind of effect rather than corrections to already-known physical magnitudes, and therefore do not challenge the existing predictions of this theory. Moreover, at a conceptual level, the diffeomorphism invariance principle and the hole argument's conclusion are not upset either, since these belong in general relativity's \emph{classical}, \emph{geometric} description of spacetime, which does not consider quantum theory at all and, therefore, must view quantum-mechanical correlations as an element foreign to its framework. In particular and furthering the previous analogy, we may say that the local comparison between $\phi_\ast[f(Q)]$ and $f(P)$ deals with changes in the components of the field, whereas the hole argument refers to (lack of) changes in the value of the fields. Thus, the \emph{classical}, \emph{geometric} theory of general relativity remains untouched from this perspective. 

Indeed, consistent with the general relativistic picture, it is not, strictly speaking, possible for the correlations we envision to be displayed by $\mathrm{d}s^2_{PP'}$ \emph{as a geometric notion}, i.e.\ as a distance between points. We interpret the appearance of quantum-mechanical correlations as an indication that there must be something amiss with the current geometric description---\emph{which leads to the futile problem of the geometric ether}---and as evidence of \emph{physical things} beyond geometry.\footnote{In fact, the consideration of quantum-mechanical ideas in themselves involves, from our point of view, the need for non-geometric physical things. In a future article, we will present quantum theory on the basis of measurement results $a_i$, and metageometric premeasurement and transition things, $\mathcal{P}(a_i)$ and $\mathcal{P}(a_j|a_i)$, familiar to human experience. We will argue that, when viewed in this manner, the theory gets rid of some of the philosophical problems that plague it (e.g.\ the geometric state vector $|\psi\rangle$ and its controversial ontology; cf.\ the geometric points $P$ and their controversial ontology).} In other words, we understand what we presently describe geometrically as ``correlations between spacetime points'' to be in fact the effect of quantum-mechanical, \emph{metageometric things}. In particular, we expect the geometric interval $\mathrm{d}s_{PP'}^2$ to result as a geometric remnant or trace of such things. 

To visualize our meaning, consider the following extension of the earlier parallel-transport analogy. Due to its homogeneity, a flat Euclidean spacetime is sterile as far as observables related to its physical geometry are concerned; a curved (pseudo-)Riemannian spacetime, on the other hand is not: the parallel transport of a field around a closed loop uncovers a previously hidden geometric observable, this spacetime's curvature. A similar relationship holds now between a \emph{vacated} general-relativistic spacetime and what we envision as a metageometric realm: whereas the former is completely homogenous concerning its elementary points (diffeomorphism symmetry) and, therefore, sterile with respect to \emph{empty} spacetime observables, the latter metageometric realm may uncover currently hidden spacetime observables as residual effects of the quantum-mechanical connection between metageometric things. 

It is through these \emph{things beyond geometry} that we aspire to achieve the earlier anticipated conceptual overthrow of the geometric ether. This is the sense, then, in which we intend the question that serves as the title of this article, and in this sense only that we intend to pursue an answer to it. In this light, one should regard points the way one would Wittgenstein's (1922) ladder: as concepts to be discarded after one ``has climbed out through them, on them, over them'' ($\S$~6.54).

\section*{References}
\begin{enumerate}

\item Auyang, S.\ Y.\ (2001). Spacetime as a fundamental and inalienable structure of fields. \emph{Studies in History and Philosophy of Modern Physics}, \emph{32}, 205--215.

\item Brans, C.\ H.\ (1999). Absolute spacetime: The twentieth century ether. \emph{General Relativity and Gravitation},
\emph{31}, 597-607, arXiv:gr-qc/9801029.

\item Butterfield, J.\ (1989). The hole truth. \emph{The British Journal for the Philosophy of Science}, \emph{40}, 1--28.

\item Drude, P.\ (1894). \emph{Physik des \"{A}thers auf elektromagnetisher Grundlage}. Stutt\-gart: Enke.

\item Earman, J.\ (1989). \emph{World enough and space-time: Absolute versus relational theories of space and time}. Cambridge, MA: MIT Press.

\item Earman, J., \& Norton, J.\ (1987). What price spacetime
substantivalism? The hole story. \emph{The British Journal for the Philosophy of Science}, \emph{38}, 515--525.

\item Einstein, A. (1952a). On the electrodynamics of moving bodies. In A.\ Einstein, H.\ Lorentz, H.\ Weyl, \& H.\ Minkowski, \emph{The principle of relativity} (pp.\ 35--65). New York: Dover (Original work published 1905).

\item Einstein, A. (1952b). The foundation of the general theory of relativity. In A.\ Einstein, H.\ Lorentz, H.\ Weyl, \& H.\ Minkowski, \emph{The principle of relativity} (pp.\ 109--164). New York: Dover (Original work published 1916).

\item Einstein, A. (1961). \emph{Relativity---The special and the general theory} (15th ed.). New York: Three Rivers Press.

\item  Einstein, A.\ (1970). Autobiographical notes. In  P.\ A.\ Schilpp (Ed.), \emph{Albert Einstein: Philosopher--Scientist} (pp.\ 1--96). LaSalle: Open Court.

\item Einstein, A.\ (1982). How I created the theory of relativity. \emph{Physics Today}, \emph{35}, 45--47
(Address delivered 14th December, 1922, Kyoto University).

\item Einstein, A.\ (1983). Ether and the theory of relativity.
In A.\ Einstein, \emph{Sidelights on relativity} (pp.\ 1--24). New York: Dover (Address delivered 5th May, 1920, University of Leyden).

\item Friedman, M. (1974). Explanation and scientific
understanding. \emph{The Journal of Philosophy}, \emph{71}, 5--19.

\item Friedman, M. (1983). \emph{Foundations of space-time theories}. Princeton: Princeton University Press.

\item Horwich, P. (1978). On the existence of time, space and
space-time. \emph{Nous}, \emph{12}, 397--419.

\item Isham, C. (1995). \emph{Lectures on quantum theory---Mathematical and structural foundations}. London: Imperial College Press.

\item Kostro, L. (2000). \emph{Einstein and the ether}. Montreal: Apeiron.

\item Kox, A.\ J.\ (1989). Hendrik Antoon Lorentz, the ether, and
the general theory of relativity. In D.\ Howard, \& J.\ Stachel
(Eds.), \emph{Einstein and the history of general relativity} (pp.\ 201--212). Boston: Birkh\"{a}user (Reprinted from \emph{Archive for History of Exact Sciences}, \emph{38} (1988)).

\item Lehto, M., Nielsen, H.\ B., \& Ninomiya, M.\ (1986a).
Pregeometric quantum lattice: A general discussion. \emph{Nuclear
Physics B}, \emph{272}, 213--227.

\item Lehto, M., Nielsen, H.\ B., \& Ninomiya, M.\ (1986b).
Diffeomorphism symmetry in simplicial quantum gravity.
\emph{Nuclear Physics B}, \emph{272}, 228--252.

\item Meschini, D., Lehto, M., \& Piilonen, J.\ (2005). Geometry, pregeometry and beyond. \emph{Studies in History and Philosophy of Modern Physics}, \emph{36}, 435--464, arXiv:gr-qc/0411053.

\item Newton, I. (1962). \emph{Mathematical principles of natural philosophy} (A.\ Motte, \& F.\ Cajori, Trans.). Berkeley: University of California Press (Original work published 1729).

\item Rynasiewicz, R.\ (1994). The lessons of the hole argument. \emph{The British Journal for the Philosophy of Science}, \emph{45}, 407--436.

\item Rynasiewicz, R.\ (1996). Is there a syntactic solution to the hole problem? \emph{Philosophy of Science}, \emph{63} (Proceedings), S55--S62.

\item Smolin, L.\ (2004, January). Atoms of space and time. \emph{Scientific American}, pp.\ 56--65.

\item Stachel, J.\ (1989). Einstein's search for general
covariance. In D.\ Howard, \& J.\ Stachel (Eds.), \emph{Einstein
and the history of general relativity} (pp.\ 63--100). Boston: Birkh\"{a}user (Original work published 1980).

\item Weyl, H.\ (1918). \emph{Raum, Zeit, Materie}. Berlin: Springer.

\item Weyl, H.\ (1949). \emph{Philosophy of mathematics and natural science}. Princeton, NJ: Princeton University Press.

\item Wheeler, J.\ A.\ (1964). Geometrodynamics and the issue of the final state. In C.\ De Witt, \& B.\ S.\ De Witt (Eds.),
\emph{Relativity, groups and topology} (pp.\ 317--520). New York:
Gordon and Breach.

\item Whittaker, E.\ T.\ (1951). \emph{A history of the theories of aether and electricity} (Vols. 1--2). London, New York: Tomash Publishers \& American Institute of Physics.

\item Wittgenstein, L.\ (1922). \emph{Tractatus logico-philosophicus} (C.\ K.\ Ogden, Trans.). London: Routledge \& Kegan.

\end{enumerate}

\end{document}